\documentclass[aps,prl,groupedaddress,superscriptaddress,twocolumn,showpacs]{revtex4}

\usepackage{graphicx}
\usepackage{amssymb}

\begin{document}

\title{Disorder-Induced Inhomogeneities of the Superconducting State
Close to the Superconductor-Insulator Transition}

\author{B. Sac\'{e}p\'{e}}
\affiliation{CEA, INAC, SPSMS-LaTEQS, 38054 Grenoble, France}
\author{C. Chapelier}
\affiliation{CEA, INAC, SPSMS-LaTEQS, 38054 Grenoble, France}
\author{T.\,I. Baturina}
\affiliation{Institute of Semiconductor Physics, 13 Lavrentjev Ave., Novosibirsk,
630090 Russia}
\author{V.\,M. Vinokur}
\affiliation{Materials Science Division, Argonne National Laboratory,
Argonne, IL 60439, USA}
\author{M.\,R. Baklanov}
\affiliation{IMEC Kapeldreef 75, B-3001 Leuven, Belgium}
\author{M. Sanquer}
\affiliation{CEA, INAC, SPSMS-LaTEQS, 38054 Grenoble, France}

\date{\today}

\begin{abstract}

Scanning tunneling spectroscopy at very low temperature on homogeneously disordered superconducting Titanium Nitride thin films reveals strong spatial inhomogeneities of the superconducting gap $\Delta$ in the density of states. Upon increasing disorder, we observe suppression of the superconducting critical temperature $T_c$ towards zero, enhancement of spatial fluctuations in $\Delta$, and growth of the $\Delta/T_c$ ratio. These findings suggest that local superconductivity survives across the disorder-driven superconductor-insulator transition.

\end{abstract}

\maketitle
A pioneering idea that in the critical region of the superconductor-insulator transition (SIT) the disorder-induced
inhomogeneous spatial structure of isolated superconducting droplets develops~\cite{Zvi94,VFGaIns96}, grew into a new paradigm~\cite{Imry}. Extensive experimental research of critically disordered superconducting films revealed a wealth of unusual and striking phenomena, including nonmonotonic temperature and magnetic field dependence of the resistance~\cite{VFGaIns96,TBJETPL,Hadacek,ShaharOvadyahu}, activated behavior of resistivity in the insulating state~\cite{Zvi94,ShaharOvadyahu,Hadacek,ZviPhysC08,TBPRL99}, nonmonotonic magnetic field dependence of the activation temperature, and the voltage threshold behavior~\cite{Shahar,TBPRL99,VVTBNat}. These features find a theoretical explanation based on the concept of disorder-induced spatial inhomogeneity in the superconducting order parameter~\cite{MaLee,DubiPRB06,Feigelman,FVB,VVTBNat}. Numerical simulations confirmed that indeed in the high-disorder regime, the homogeneously disordered superconducting film breaks up into superconducting islands separated by an insulating sea~\cite{Ghosal,DubiNat}. At the same time the direct observation of superconducting islands near the SIT justifying the fundamental but yet hypothetical concept of the disorder-induced granularity on the firm experimental foundation was still lacking.

In this Letter, we report on the combined low temperature Scanning Tunneling Spectroscopy (STS) and transport measurements performed on thin Titanium Nitride films on approach to the SIT. The local tunneling density of states (LDOS) measured at 50\,mK reveals
disorder-induced spatial fluctuations of the superconducting gap, $\Delta$, with both, standard deviation $\sigma$ to the average gap
and the gap to the critical temperature ratios, $\sigma/\overline{\Delta}$ and $\overline{\Delta} / T_c$, respectively, increasing towards the transition.

Our samples were thin TiN films synthesized by atomic layer chemical vapor deposition onto a Si/SiO$_2$ substrate. TiN1 was a 3.6\,nm thick film deposited at 400$^{\circ}$C while TiN2 and TiN3 were 5.0\,nm thick films deposited at 350$^{\circ}$C.
TiN3 was then slightly plasma etched in order to reduce its thickness. Electron transmission images revealed that the films comprise of the densely-packed crystallites with a typical size of 4 to 6\,nm. The samples were patterned into the Hall bridges using conventional UV lithography and plasma etching. It is worth noticing that identically fabricated TiN films undergo the disorder- and magnetic field-driven SIT~\cite{TBJETPL,TBPRL99}. Transport measurements and STS were carried out during the same run in a STM attached to a dilution refrigerator. The STM Pt/Ir tip was aligned above one of the free $500 \times 500\,\mu$m$^2$ contact pads of the Hall bridge. In order to probe the LDOS, the differential conductance of the tunnel junction, G(V), was measured by a lock-in amplifier technique with an alternative  voltage of 10\,$\mu$V added to the direct bias voltage. Resistances were measured by acquiring both voltage and current with a low frequency lock-in amplifier technique in a four wires configuration.

Shown in Fig.\,1a are the temperature dependences of sheet resistances, $R_{\square}$, for the three samples, which room
temperature sheet resistances are listed into Table I. All three curves show nonmonotonic behavior with a pronounced maximum
preceding the superconducting transition. Such a behavior can be described within the framework of the theory of quantum corrections for quasi-two-dimensional disordered superconductors. Making use of theoretical expressions for corrections~\cite{AAreview,LarVar,AVR1983}, we can find the superconducting temperature $T_c$~\cite{Benj} (see Table I).
The calculated fits are presented by solid lines in Fig.\,1a. The perfect description of the experimental curves achieved by means of formulas derived for homogeneously disordered systems, shows that relying on the macroscopic parameter R(T), we could have expected that our homogeneously disordered films possess a spatially uniform superconducting state.  However, using the data obtained by virtue of the local STM probe, we reveal that due to mesoscopic fluctuations of disorder, superconducting properties show inhomogeneities on mesoscopic spatial scales. Typical local tunneling data are shown in Fig.\,1b.
In contrast with earlier macroscopic tunneling studies performed with lithographied junctions on Bismuth films \cite{Valles}, our local STM measurements systematically showed fully gapped shapes for any degree of disorder indicating the absence of quasiparticles at the Fermi level. These BCS-like spectra at low energy allowed us to extract $\Delta$ values which were significantly reduced as compared to $\Delta^{bulk} = 730 \:\mu$eV in bulk TiN~\cite{Escoffier}. Yet, spectra taken at different positions on the surface give different values of $\Delta$. This evidences that the superconducting state is spatially inhomogeneous.

\begin{figure}[tb]
\includegraphics[width=80mm]{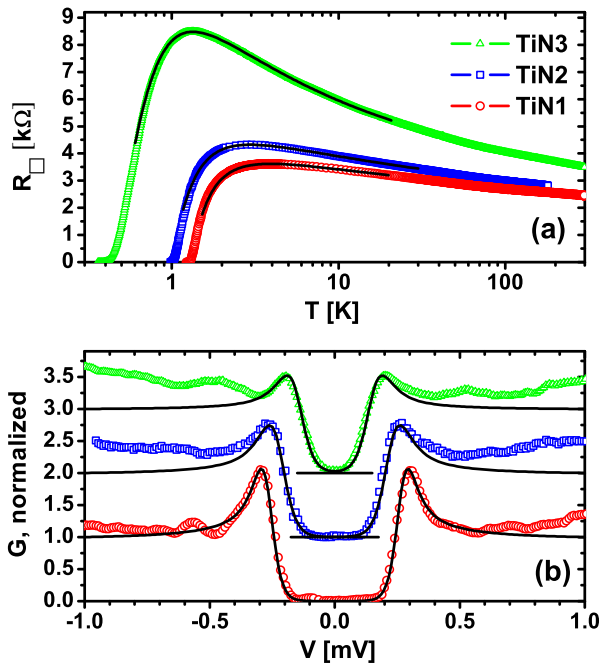}
\caption{\label{Figure1} (a)~Sheet resistance $R_{\square}$ versus temperature for three samples. The solid lines are fits according to localization-interaction and superconducting fluctuations corrections. The legend of panel (a) describes the two panels. (b)~Normalized differential tunneling conductance measured at $T = 50$\,mK (dots). Spectra are shifted for clarity. The BCS fits (solid lines) were calculated with the following parameters: TiN1 - $\Delta = 260\:\mu$eV and an effective temperature 
$T_{eff} = 0.25$\,K; TiN2 - $\Delta = 225\:\mu$eV, $T_{eff} = 0.32$\,K; TiN3 - $\Delta = 154\:\mu$eV, $T_{eff} = 0.35$\,K.}
\end{figure}

\begin{table}[h!]
\caption{\label{tab}Sample characteristics: $R_{300}$ - resistance per square at 300\,K. $T_c$ - critical temperature determined from
the quantum correction fits. $\overline{\Delta}$ - average value of the superconducting gap. $\sigma$ - standard deviation of the
superconducting gap.}
\begin{ruledtabular}
\begin{tabular}{cccccccc}
 & $R_{300}$ &  $T_c$ &  $\overline{\Delta}$ & $\sigma$& $\sigma /\overline{\Delta}$ &$\overline{\Delta} / k_BT_{c}$  \\
 & k$\Omega $ &   K   &  $\mu$eV             &  $\mu$eV&                             &     \\
\hline
TiN1 & 2.45 & 1.3  &  265 & 11 & 0.04 & 2.37\\
TiN2 & 2.7  & 1.0  &  220 & 13 & 0.06 & 2.55\\
TiN3 & 3.5  & 0.45 &  160 &  - &  -   & 4.13\\
\end{tabular}
\end{ruledtabular}
\end{table}

\begin{figure}[tb]
\includegraphics[width=80mm]{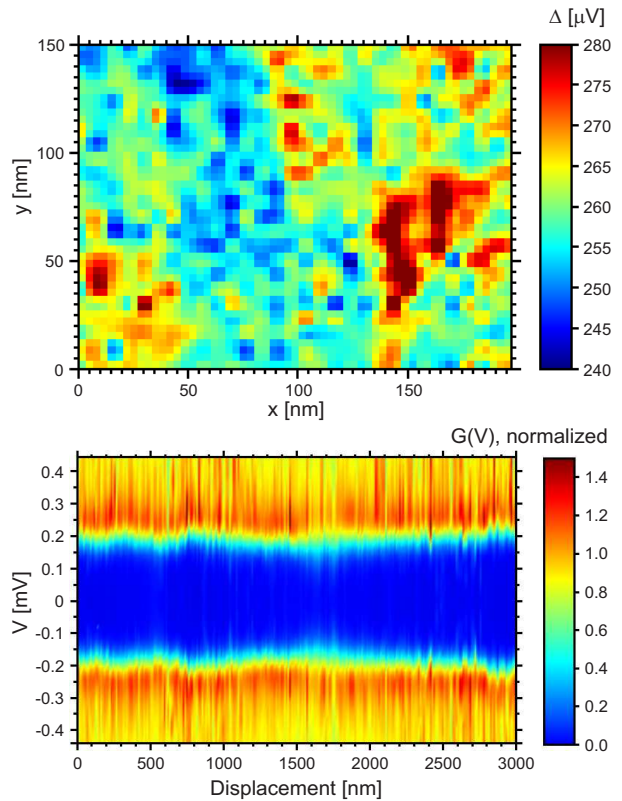}
\caption{\label{Figure2}
Top: The colour map of spatial fluctuations of $\Delta$ on TiN1. Inhomogeneities of the superconducting properties show up on a scale of a few tens of nanometers.
Bottom: Spectra measured along a straight line on TiN2. The BCS-like LDOS fluctuates symmetrically around the Fermi level.}
\end{figure}

The top panel of Fig.\,2 presents the spatial map of superconducting gap $\Delta$ measured over the $200 \times 150$\,nm$^2$ area with a pitch of 3.3\,nm. The characteristic scale of the inhomogeneity is estimated as a few tens of nanometers. The measured 2700 spectral gap amplitudes give a gaussian distribution with an average value $\overline{\Delta} = 265 \:\mu$eV and a standard deviation $ \sigma = 11 \:\mu$eV. The spatial fluctuations are straightforwardly seen in the bottom panel of Fig.\,2 displaying a color map of 256 LDOS spectra measured on TiN2 every 11.7 nm along a straight line. Note, that spectra are symmetric with respect to voltage direction. We find $\overline\Delta = 220 \:\mu$eV and $\sigma = 13 \:\mu$eV. In TiN3, probably because of the plasma etching, the surface did not suit for spatially regular STS measurements. Therefore we performed point-contact spectroscopy, the STM tip gently touching the surface sample.  From the 30 BCS-like spectra we obtained the magnitudes of the gap scattered in the interval from 125 to 215\,$\mu$eV, with an average $\overline{\Delta} = 160 \:\mu$eV.  For the three samples, every spectra measured in the scanning window of 3 x 3 $\:\mu m ^2$ displayed gaps consistent with the evaluated distribution. All the data are summarized in Table I. They show an unusually large (as compared to the BCS-predicted 1.76 value) $\overline{\Delta}/k_BT_{c}$ ratio and an increasing relative standard deviation $\sigma/\overline{\Delta}$ with disorder.

When dealing with STM and spatially resolved superconducting inhomogeneities, the question of the tunneling junction quality cannot be eluded. However, the lack of states at the Fermi level in our measurements is a strong indication of good tunneling conditions. For instance, for tunneling from a damaged metallic surface weakly coupled to the film with a proximity-induced superconducting LDOS, one would have rather observed small $\Delta /k_BT_c$ ratios and a significant amount of states at the Fermi level. These fully gapped spectra also rule out any spurious inelastic scattering of the tunneling carriers. Moreover the diffusive length probed by them in the mV range is about the thickness of the film itself which makes questionable any distinction between the physical properties of the surface and of the film. Finally, we did not observe any dependence of these spectra upon the tip-sample distance which varied only slightly for TiN1 and TiN2 when the tunneling current setpoint was changed over orders of magnitude. All this makes us confident that the discussed superconducting inhomogeneities are of the intrinsic origin.

The appearance of spatial inhomogeneities of the order parameter was indeed predicted by theoretical calculations \cite{MaLee, Feigelman} and by numerical simulations \cite{Ghosal} as a result of interplay between superconductivity and Anderson localization in a strongly disordered superconductor. The recent work~\cite{Skvortsov} showed that even in the weakly disordered systems far from the localization threshold, the mesoscopic fluctuations of the conductance near the point of the Coulomb suppression of superconductivity~\cite{Finkelstein} also give rise to spatial inhomogeneities of the order parameter. We like to emphasize that in all the above works it was the short range, i.e. atomic scale, disorder that caused variations in the physical characteristics of the system on the mesoscopic scale and thus drove it into an inhomogeneous superconducting state.
Numerical simulations~\cite{Ghosal} demonstrated further that increasing disorder washes out the BCS singularities by pushing states at higher energies. We indeed observe smearing out of these coherence peaks. However, we can well describe this effect with BCS formulas with the effective temperature $T_{eff}$ which is higher than the expected 250 mK set-up resolution \cite{Sacepe}. Note at this point that BCS expressions do not describe the observed upturn in the LDOS above the gap (see Fig.\,1b) which is likely due to Coulomb interactions \cite{AAreview} not considered in ref. \cite{Ghosal}. We conjecture that $T_{eff}$ is a consequence of the spatial modulation of the superconducting gap. The regions with the reduced $\Delta$ act as traps for normal excitations. Tunneling of a quasiparticle between the regions with the reduced gap gives rise to a broadening, $\Gamma$, of the quasiparticle energy levels. This broadening is equivalent to some effective temperature, $T_{eff}\simeq\Gamma/k_B$, and can be estimated as $\Gamma\simeq \hbar^2 I/(2m^{\ast} L^2)$, where $L$ is the characteristic spatial scale of the modulation of the superconducting gap, $I\simeq\exp(-L/\xi)$ is the tunneling integral, and $m^{\ast}$ is the quasiparticle mass. Taking for a crude evaluation $L\simeq\xi\approx 10$\,nm (the latter is derived from the experimental value of the upper critical field $B_{\mathrm{c2}}(0)=2.8$\,T~\cite{PhysBTB}) and $m^{\ast} \sim 2m_e$~\cite{Greeks}, we arrive at $T_{eff}\approx 1$\,K, which is amazingly close to experimental findings given very approximate character of our estimate. Note that the effect of smearing of the BCS peaks in the average density of states of inhomogeneous superconductors was discussed in the classical paper by Larkin and Ovchinnikov~\cite{LarkinOvchin71n2}.

\begin{figure}[tb]
 \includegraphics[width=80mm]{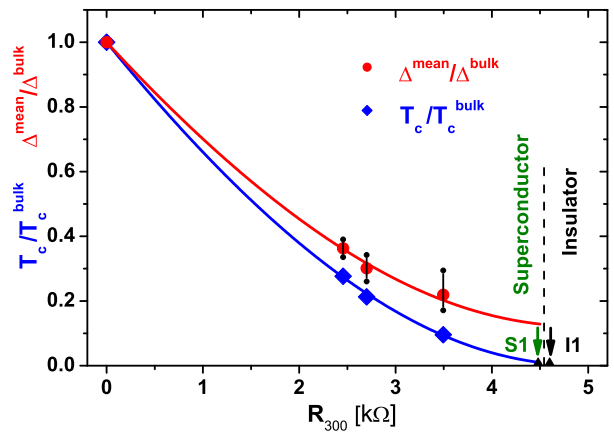}
 \caption{\label{Figure3}
Critical temperature $T_c$ and mean spectral gap $\overline\Delta$ at $T = 50$\,mK divided by their bulk value versus the room temperature sheet resistance. Black bars represent the measured gap ranges for each sample ($\sim 4 \sigma$ for TiN1 and TiN2). Blue solid line is a fit according to Finkelstein's model~\cite{Finkelstein}. Red solid line is a guide to the eye. Black triangles are the room temperature sheet resistances for the \textquotedblleft last\textquotedblright superconducting sample (S1 - $R_{300} = 4.48$\,k$\Omega$) and \textquotedblleft first\textquotedblright insulating sample (I1 - $R_{300} = 4.60$\,k$\Omega$) from the Ref.~~\cite{TBPRL99}.
}
\end{figure}

Now we are in a position to discuss the evolution of the superconducting properties of the TiN films with increasing disorder, measured by $R_{300}$, the room temperature sheet resistance. Fig.\,3 presents $\overline\Delta/\Delta^{bulk}$ and $T_c/T_c^{bulk}$ versus $R_{300}$ for the samples under this study. $\Delta^{bulk}= 0.73\,meV$ and $T_c^{bulk} = 4.7\,K$  were measured on a much less disordered thick film with $R_{300}= 27\, \Omega$ \cite{Escoffier}. In addition, the data on the sample S1, with $R_{300} = 4.48$\,k$\Omega$, and on the sample I1, with $R_{300} = 4.60$\,k$\Omega$ from the Ref.~\cite{TBPRL99} are shown. 

Important inferences can be deduced from this figure:

(i) Increasing disorder suppresses $T_c$ towards its complete vanish.
This behavior is well fitted within Finkel'stein's model which describes disorder-enhanced Coulomb interaction in \textit{homogeneously} disordered thin films \cite{Finkelstein}. In this model, the reduction of $T_c$ follows the relation : $\ln\left(\frac{T_c}{T_c^{bulk}}\right)=\gamma+\frac{1}{\sqrt{2r}}\ln\left(\frac{1/ \gamma + r/4 - \sqrt{r/2}}{1/ \gamma + r/4 + \sqrt{r/2}}\right)$, where $r=R_\Box e^2/(2\pi ^2 \hbar)$ and $\gamma$ is the only fitting parameter. We found $\gamma \simeq 6.2$ consistent with the previously reported data on TiN ($\gamma \simeq 6.8$)\cite{Hadacek}. Note that the samples S1 and I1 from Ref.~\cite{TBPRL99} are identically fabricated TiN films \textit{closest} to the SIT from the superconducting and insulating sides respectively.

(ii) The variance in the gap magnitude increases with disorder. The enhancement of spatial fluctuations leads to a strongly inhomogeneous superconducting state when approaching the SIT. This observation supports numerical calculations~\cite{Ghosal}. It further agrees with the conclusion of~\cite{Skvortsov} that the mesoscopic fluctuations of the Cooper attraction constant in Finkel'stein's theory yield strong spatial fluctuations of the local order parameter close to the critical disorder.

(iii) The suppression of $T_c$ advances that of $\overline\Delta$, yielding anomalously large $\overline\Delta / k_BT_{c}$ ratios
(see Table 1) as compared to the bulk TiN value $\Delta^{bulk} / k_B T_c^{bulk}=1.8$ \cite{Escoffier}. This can indicate, should the tendency displayed on Fig.\,3 persisted till the full suppression of $T_c$, that despite vanishing of the phase stiffness at the critical resistance, the average superconducting gap remains finite. 
Extrapolating the red line over the insulating side of SIT (Fig.\,3), one then could have concluded that even in the insulating phase where $T_c = 0$, the average superconducting gap still survives. Thus the observed trend of the faster decay in $T_c$ than in $\overline\Delta$, together with the inhomogeneous state, offers a strong support to hypothesis that the so-called \textquotedblleft homogeneously disordered \textquotedblright  superconducting films near SIT~\cite{Zvi94,VFGaIns96,Ghosal,TBPRL99} can in fact be viewed as granular-like superconducting structure or Josephson phase suggested by Imry \textit{et al}~\cite{Imry}. This, in its turn, implies that the bosonic mechanism of the superconductor-to-insulator transition~\cite{Fisher} becomes relevant. However, an alternative theory building a superconducting state based on fractal wavefunctions close to the Anderson transition has been recently proposed \cite{Feigelman}. Consistently with our data, the authors predict likewise an inhomogeneous superconductor with strong $\Delta / k_BT_{c}$ ratios and a gapped insulator with no coherence peaks due to pairing of electrons on localized states.

In conclusion, our results demonstrate that the behavior of \textit{global} characteristics, such as $R_{\square}(T)$ and the suppression of $T_c$ by increasing disorder in the TiN films near the superconductor-insulator transition are described perfectly well by theories developed for homogeneously disordered films. At the same time the measurements of the \textit{local} tunneling density of states clearly indicate the existence of the inhomogeneous superconducting state.
Moreover, the data show that LDOS is likely to remain gapped beyond the SIT in the insulating phase. This suggests the presence of the regions with the finite superconducting gap in the insulating side of this transition. Several issues related to our results on local tunneling measurements call for further research.  First, the non-constant LDOS above the gap seen in the lower panel of Fig.\,1 requires thorough analysis taking into account Coulomb interactions. Another important question is the role of spatial modulation of the gap.  While its effect on the \textit{global} density of states averaged over the whole system was discussed in~\cite{LarkinOvchin71n2}, the consistent theoretical description of the \textit{local} DOS is still lacking.

\begin{acknowledgments}
We are grateful to M. Feigel'man for stimulating comments on this work and like to thank B. Altshuler, A. Finkelstein, \O. Fischer, M. Houzet, Y. Imry, V. Mineev, and  A. Varlamov for fruitful discussions. This research is supported by the Program ``Quantum macrophysics'' of the Russian Academy of Sciences, the Russian Foundation for Basic Research (Grant No. 06-02-16704), and the U.S. Department of Energy Office of Science under the Contract No. DE-AC02-06CH11357.
\end{acknowledgments}

\end{document}